\documentclass[twoside,slac_one,a4paper]{revtex4}
\usepackage{graphicx}
\usepackage{fancyhdr}
\usepackage{amsmath} 
\usepackage{bm}
\usepackage{amsxtra}
\usepackage{amssymb}
\usepackage{amsthm}
\usepackage{latexsym}
\usepackage{lscape}

\begin{document}

\title{Daily Modulation of the Dark Matter Signal in Crystalline Detectors}

%

\author{Nassim Bozorgnia}
\affiliation{Department of Physics and Astronomy, UCLA, 475 Portola Plaza, Los Angeles, CA 90095, USA}

\begin{abstract}
The channeling effect in crystals refers to the orientation dependence of charged ion penetration in crystals. In direct dark matter crystalline detectors, a channeled ion recoiling after a collision with a WIMP gives all its energy to electrons. Thus channeling increases the ionization or scintillation signal expected from a WIMP. Channeling is a directional effect which depends on the velocity distribution of WIMPs in the dark halo of our Galaxy and could lead to a daily modulation of the signal. I will present estimates of the expected amplitude of the daily modulation in direct dark matter detectors, both due to channeling and just due to the rotational velocity of the Earth around itself.
\end{abstract}

\maketitle

\thispagestyle{fancy}


\section{Introduction}
Understanding ion channeling and blocking in crystals is important in the interpretation of direct dark matter experiments measuring ionization or scintillation signals. In the ``channeling effect'' ions incident upon a crystal along symmetry axes and planes suffer a series of small-angle scatterings that maintain them in the open ``channels''  between the rows or planes of lattice atoms and thus lose their energy predominantly to electrons (while non-channeled ions transfer most of their energy to lattice nuclei). The ``blocking effect''  consists in a reduction of the flux of ions originating in lattice sites along symmetry axes and planes due to the shadowing effect of the lattice atoms directly in front of the emitting lattice site~\cite{Gemmell:1974ub}.

The  potential importance of the channeling effect for direct  dark matter detection was first pointed out for NaI (Tl) by Drobyshevski~\cite{Drobyshevski:2007zj} and subsequently by the DAMA collaboration~\cite{Bernabei:2007hw}. When  Na or I ions recoiling after a collision with a  dark matter WIMP (Weakly Interacting Massive Particle) move along crystal axes and planes, their quenching factor is approximately $Q=1$ instead of $Q_{\rm Na}=0.3$ and $Q_{\rm I}=0.09$, since they give their energy to electrons. The DAMA collaboration estimated the fraction of channeled recoils and found it to be large for low recoiling energies in the keV range. This effect considerably shifts the regions in cross section versus mass of acceptable WIMP models in agreement with the DAMA data towards lower WIMP masses. As a consequence, the comparison between the DAMA results and the null results of other experiments is affected too~\cite{Savage:2010}.

In 2008, Avignone, Creswick, and Nussinov~\cite{Avignone:2008cw} suggested that ion channeling in NaI crystals could give rise to a daily modulation of the dark matter signal. 
Such a modulation of the rate due to channeling is expected to occur at some level because the ``WIMP wind'' arrives to Earth on average from a particular direction fixed to the Galaxy. Assuming that the dark matter halo is on average at rest with respect to the Galaxy, this is the direction towards which the Earth moves with respect to the Galaxy. Earth's daily rotation naturally changes the direction of the ``WIMP wind'' with respect to the crystal axes, thus changing the amount of recoiling ions that are channeled vs.~non-channeled. This amounts to a daily modulation of the dark matter signal detectable via scintillation or ionization. In addition to the daily modulation due to channeling, there exists a daily modulation in the usual (i.e. non-channeled) signal rate which is due purely to the change of WIMP kinetic energy in the lab frame as the Earth rotates around itself. This means that the number and energy of WIMPs above threshold in the lab frame changes during a day. If measured, the daily modulation due to channeling or in the usual rate would be a background free dark matter signature.

My collaborators, Graciela Gelmini and Paolo Gondolo, and I~\cite{BGG} have evaluated the upper bounds on the channeling fractions for different crystalline detectors used in dark matter experiments through analytical means. To calculate the amplitude of the daily modulation in NaI crystals, we compute the event rate during a day using the channeling fractions and the actual differential recoil spectrum and taking into account channeled and non-channeled recoils. We also examine the possibility that such a daily modulation might be observable in the data accumulated by the DAMA collaboration.

\section{Channeling Fractions}
Our calculation of the channeling fractions as function of recoil energy is based on  the classical analytic models which started to be developed in the 1960's and 70's. In particular we use Lindhard's model~\cite{Lindhard:1965}, supplemented by the planar channel models of Morgan and Van Vliet~\cite{Morgan-VanVliet} and the work of Hobler~\cite{Hobler} on low energy channeling.
In these models the rows and planes of lattice atoms are replaced by continuum strings and planes, in which the screened Thomas-Fermi potential is averaged over a direction parallel to a row or a plane. The continuous potential, $U$ is considered to be uniformly smeared along the row or plane of atoms, which is a good approximation if the propagating ion interacts by a correlated series of many consecutive glancing collisions with lattice atoms in the row or plane. Just one row or plane is considered in this model. Lindhard proved that for an ion propagating with kinetic energy $E$, and for small  angle $\phi$  between the ion's trajectory and  the atomic  row (or plane) in the direction perpendicular to  the row (or plane), the  so called ``transverse energy'', $E_{\perp} = E \sin^2\phi + U\simeq E \phi^2 + U$ is conserved.

The conservation of the transverse energy provides a definition of the minimum distance of approach to the string (or plane) of atoms, $\rho_{\rm min}$, at which the trajectory of the ion makes  a zero angle with the string (or plane), and also of the angle $\psi$ at which the ion exits from the string (or plane), i.e. far away from it where $U \simeq0$. Channeling requires that $\rho_{\rm min} > \rho_c(E)$, where $\rho_c(E)$ is the smallest possible minimum distance of approach of the propagating ion with the row (or plane) for a given energy $E$. This amounts to $\psi < \psi_{c}(E)$, where $\psi_{c}(E)$  is the maximum angle  the ion can make with the string far  away from it (i.e. in the middle of the channel) if the ion is channeled.

\subsection{Channeling of Incoming Particles}
The channeling of ions in a crystal depends not only on the angle their initial trajectory makes with  strings or planes in the crystal, but also on their initial position. Ions which start their motion close to the center of a channel,  far from a string or plane, where they make an angle $\psi$, are channeled if the angle  is smaller than a critical angle (as explain earlier) and are not  channeled otherwise. Particles which start their motion in the middle of a channel (as opposed to  a lattice site) must be incident upon the crystal.

No data or simulations of Na and I ions propagating in a NaI crystal is available at low energies. We show that  to a good approximation we can use analytic calculations and reproduce the channeling fraction in NaI presented in Ref.~\cite{Bernabei:2007hw}. Temperature corrections are neglected and we use a static lattice (similar to the approach of Ref.~\cite{Bernabei:2007hw}). For an incident angle $\psi$ with respect to each of the channels and an ion energy $E$, the fraction of channeled incident ions for axial and planar  channels  is 1 if $\psi$ is smaller than the critical angle for the corresponding channel and zero otherwise.
\begin{figure}
\begin{center}
   \includegraphics[height=190pt]{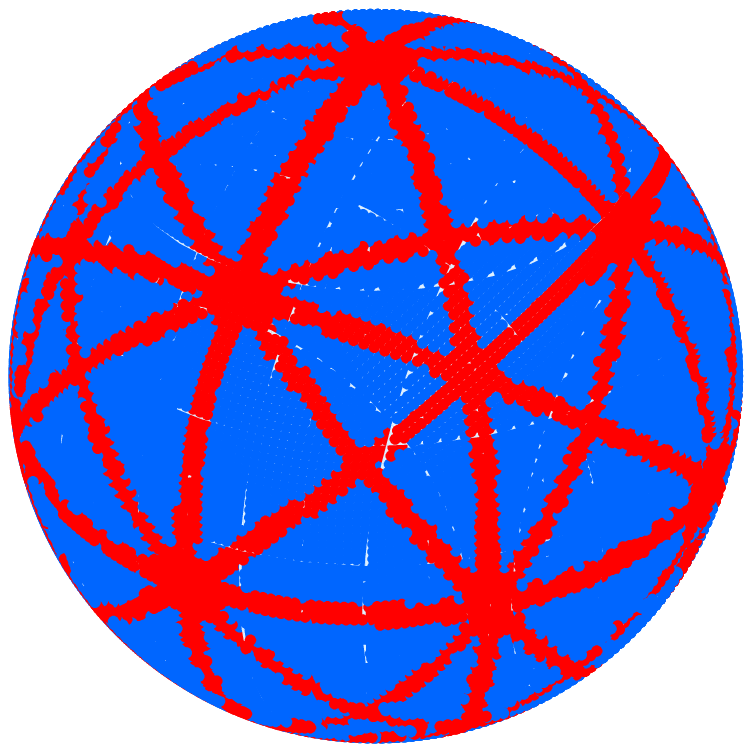}
  \includegraphics[height=160pt]{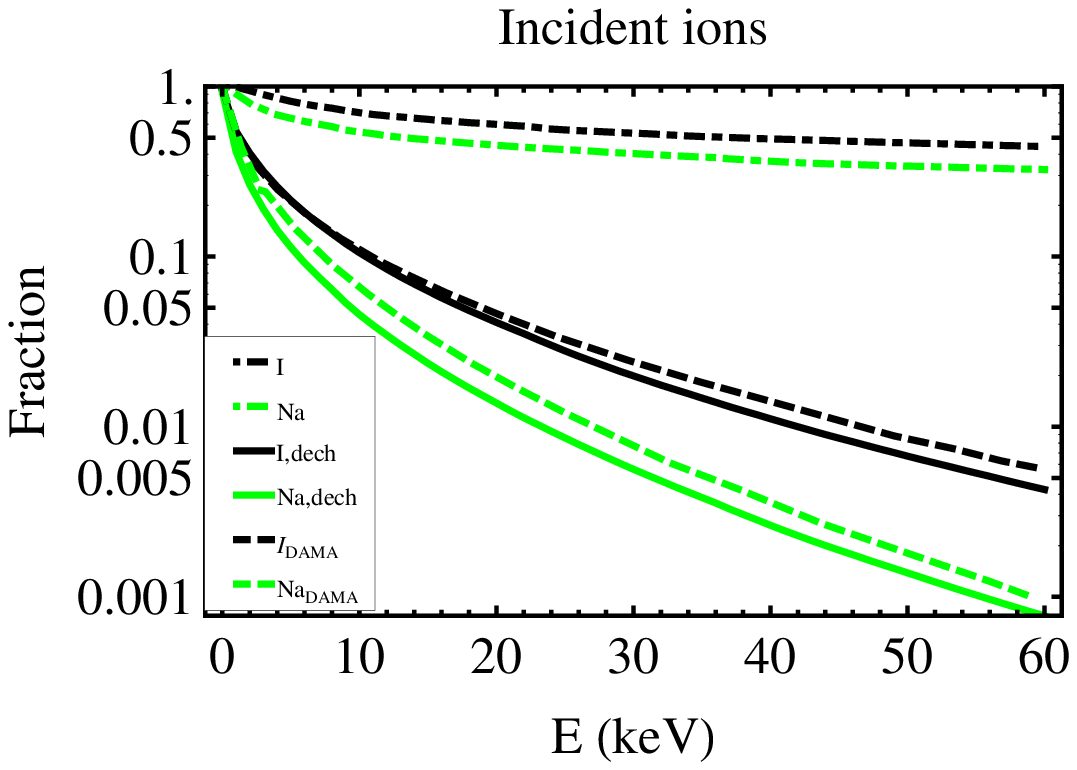}
 \caption{(a) Channeling fraction for a 50 keV Na ion in different directions plotted on a sphere using HEALPix:  probability equal to one in red, and probability equal to zero in blue. (b) Fraction of channeled incident  I (black) and Na (green/gray) ions as a function of their incident energy $E$ with the static lattice without  (dot dashed lines) and with (solid lines) dechanneling due to interactions with Tl impurities.  The results of DAMA are also included (dashed lines).}
  \label{DAMA}
\end{center}
\end{figure}

To find the total fraction of channeled incident nuclei, we integrate the direction-dependent channeling fraction over the incident direction. The integral cannot be solved analytically, so we integrate numerically by performing a Riemann sum once the sphere of directions has been divided using the Hierarchical Equal Area iso-Latitude Pixelization (HEALPix)~\cite{HEALPix:2005}. HEALPix uses an algorithm to deal with the pixelization of data on a sphere, and it is useful to compute integrals over direction by dividing the surface of a sphere into many pixels, computing the integrand at each pixel (i.e. each direction, see Fig.~\ref{DAMA}.a), and finally summing up the values of all pixels over the sphere.

A  channeled ion can be pushed out of a channel by  an interaction with an  impurity such as the atoms of Tl in NaI (Tl). Here we simply assume that if a channeled ion interacts with a Tl atom it becomes dechanneled and thus it does not contribute to the fully channeled fraction any longer. We thus neglect  the possibility that after the interaction the ion may reenter into a channel, either the same or another. Figure \ref{DAMA}.a shows the axial and planar channels of the NaI crystal for  incoming  Na ions with an energy of 50 keV. We include here only the channels with lower crystallographic indices, i.e. 100, 110 and 111, which provide the dominant contribution to the channeling fraction, as is also done in Ref.~\cite{Bernabei:2007hw}. The fraction of channeled incident Na or I ions as a function of their incident energy $E$ is shown in  Fig.~\ref{DAMA}.b. For comparison, Fig.~\ref{DAMA}.b also shows the channeling fraction obtained by DAMA. Good agreement with the channeling fractions of DAMA is achieved only when dechanneling due to the interaction with Tl impurities is included.

\subsection{Channeling of Recoiling Lattice Nuclei}
The recoiling nuclei start initially from lattice sites (or very close to them), thus blocking effects  are important. In fact, as argued originally by Lindhard~\cite{Lindhard:1965}, in a perfect lattice and in the absence of energy-loss processes the probability that a particle starting from a lattice site is channeled would be zero. The argument uses statistical mechanics in which the probability of particle paths related by time-reversal is the same. Thus the probability of an incoming ion to have a particular path within the crystal is the same as the probability of the same ion to move backwards along the same path~\cite{Gemmell:1974ub}. This is what Lindhard called the ``Rule of Reversibility". Using this rule, since the probability of an incoming channeled ion to get very close to a lattice site is zero, the probability of the same ion to move in the time-reversed path,  starting at a nuclear site and ending inside a channel is also zero. However, any departure of  the actual lattice from a perfect lattice, for example due to vibrations of the atoms in the lattice, would violate the conditions of this argument and allow for some of the recoiling lattice nuclei to be channeled, as already understood in the 70's~\cite{Komaki:1970}.

In  our model, a recoiling ion is channeled if the collision ion-WIMP happens at a distance large enough from  the string  or plane to which the ion belongs. Namely, channeling happens if the initial position of the recoiling motion is $\rho_i > \rho_{i,\rm min}$. We define $E_{\perp}$ in terms of the initial recoil energy $E$ of the propagating ion, the angle of the initial recoil momentum  with respect to the particular string or plane of atoms $\phi_i$, and the initial position $\rho_i$. Since the potential $U(\rho)$  decreases monotonically with increasing $\rho$, $U(\rho_{\rm min}) < U(\rho_c(E))$, and the condition
\begin{equation}
E_{\perp}(E,\phi_i,\rho_i)= U(\rho_{\rm min}) < U(\rho_c),
\label{ChanCond-CA}
\end{equation}
i.e. $E_{\perp}(E,\phi_i,\rho_{i,\rm min})= U(\rho_c(E))$ defines the distance $\rho_{i,\rm min}$ in terms of $\rho_c$.

We take the initial distance distribution, $\rho_i$ of the colliding atom to be a Gaussian with  a  one dimensional rms vibration amplitude $u_1$ (using the Debye model), and we obtain the probability of channeling for each individual channel  by integrating the Gaussian between the minimum initial distance and infinity (a good approximation to the radius of the channel). The dependence of these probabilities on the critical distances enter in the argument of an exponential (for axial channels) or an erfc function (for planar channels). In order to obtain the total geometric channeling fraction we sum over all the individual channels we consider. The integral over initial directions is computed using HEALPix.

\begin{figure}
\begin{center}
  \includegraphics[height=180pt]{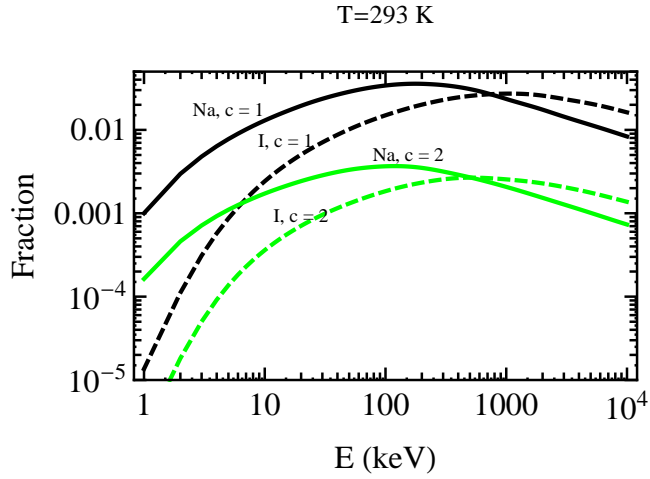}
  \caption{Upper bounds to the channeling fraction of recoiling Na (solid lines) and I (dashed lines) ions in NaI for $c=1$ (black) and $c=2$ (green/gray) cases at T=293 K.}
  \label{FracNaI-Final}
\end{center}
\end{figure}

Figure \ref{FracNaI-Final} shows what we consider to be our main predictions for the range expected as an upper limit to the channeling fraction in NaI at 239 K. The parameter $c$ mentioned in the figure is a number that we expect to be between 1 and 2, which regulates the importance of temperature corrections. The channeling fractions are typically smaller for larger values of $c$. Notice that we have not included any dechanneling effects due to impurities or dopants, which should also decrease the channeling fractions. Using the fractions in Fig.~\ref{FracNaI-Final}, the difference in the regions of cross section versus mass of acceptable WIMP models corresponding to the DAMA annual modulation signal differs only at 7$\sigma$ if channeling is included or not included~\cite{Savage:2010}.
In the absence of channeling, the DAMA region could be compatible with the CoGeNT region for light WIMPs~\cite{Hooper}.

\begin{figure}
\begin{center}
  \includegraphics[height=160pt]{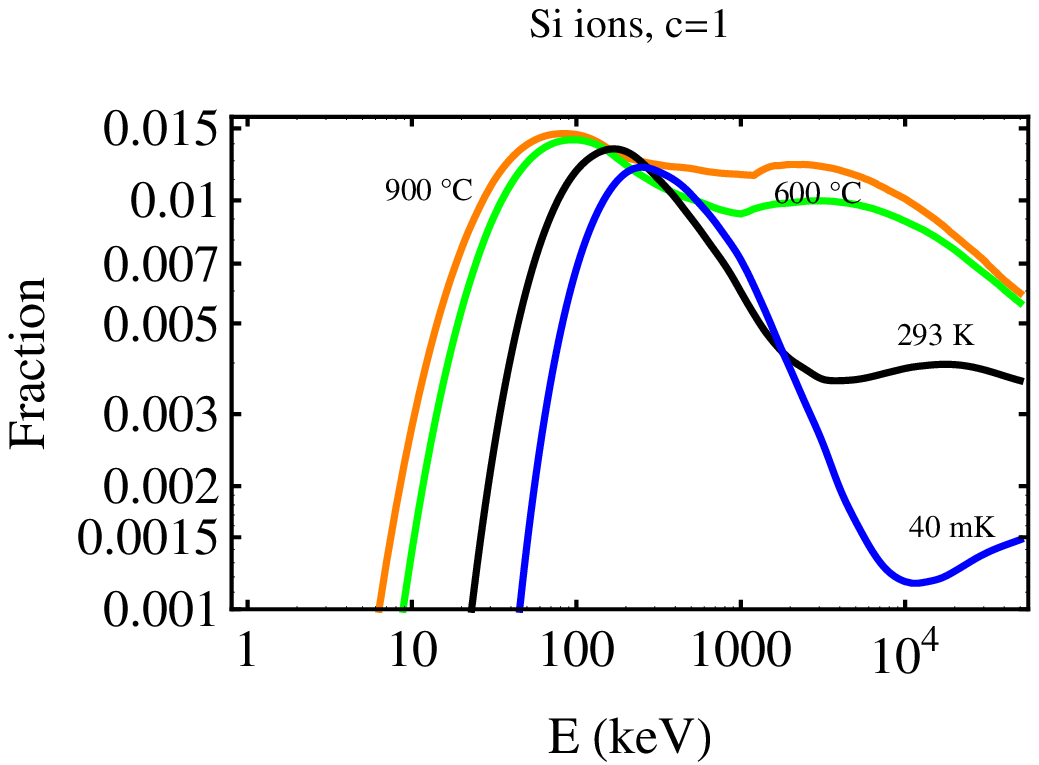}
  \includegraphics[height=160pt]{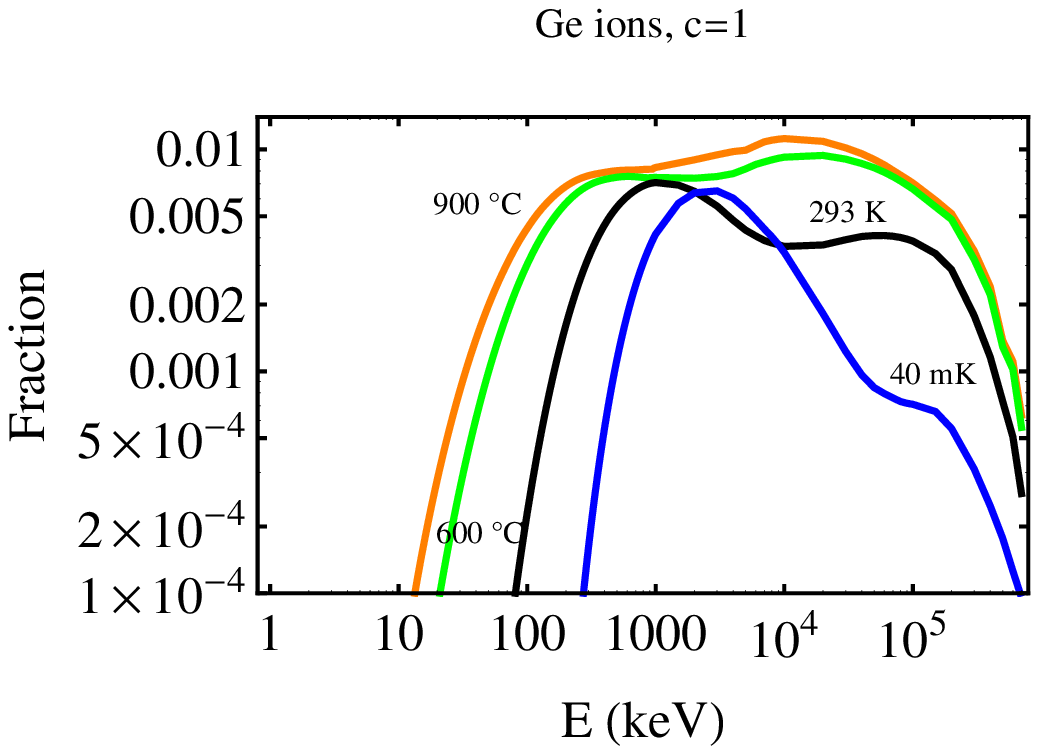}\\
  \caption{Channeling fractions of (a) Si  and (b) Ge recoils in a Si and a Ge crystal respectively, as a function of the ion energy for  temperatures $T=900~^\circ$C (orange or medium gray), 600 $^\circ$C (green or light gray), 293 K (black), and 44 mK (blue or dark gray)  in the approximation of $c=1$.}
  \label{FracSiG-DiffT-c1}
\end{center}
\end{figure}

\begin{figure}
\begin{center}
  \includegraphics[height=160pt]{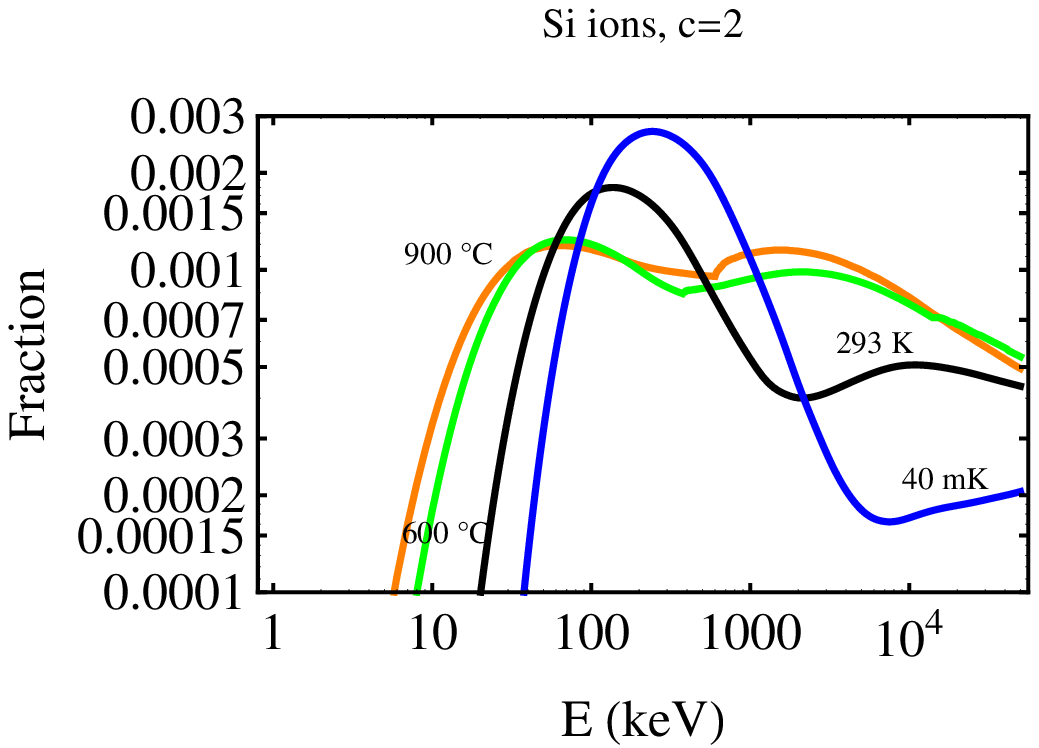}
  \includegraphics[height=160pt]{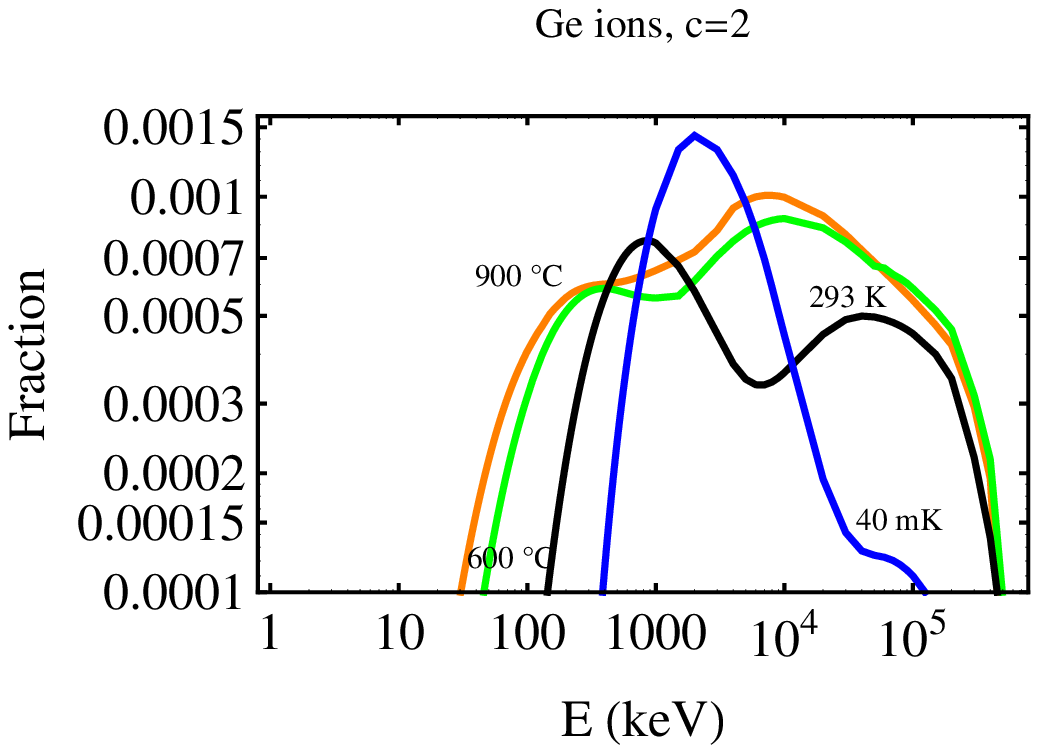}\\
  \caption{Same as Fig.~\ref{FracSiG-DiffT-c1} but for $c=2$.}
  \label{FracSiG-DiffT-c2}
\end{center}
\end{figure}

Our results for the total geometric channeling fraction for Si ions propagating in a Si crystal  and  Ge ions propagating in a Ge crystal at different temperatures are shown in Figs.~\ref{FracSiG-DiffT-c1} and \ref{FracSiG-DiffT-c2} for the two cases of $c=1$ and $c=2$, respectively. The channeling fractions we obtain are strongly temperature dependent. As the temperature increases, the probability of finding atoms far from their equilibrium lattice sites increases, which increases the channeling fractions, but the critical distances $\rho_c$ become larger which decreases the channeling fractions.
%

\section{Angular Distribution of Recoil Directions due to WIMPs}
Consider the WIMP-nucleus elastic collision for a WIMP of mass $m$ and a nucleus of mass $M$. The 3-dimensional ``Radon transform'' of the WIMP velocity distribution can be used to define the differential recoil spectrum as a function of the recoil momentum $\vec{\bf q}$~\cite{Gondolo:2002}
\begin{equation}
\frac{dR}{ dE_R~d\Omega_q}= \frac{\rho \sigma_0 S(q)}{4\pi m \mu^2} \hat{f}_{\rm lab}\!\left( \frac{q}{2\mu}, \hat{\bf q} \right) ,
\label{eq: rate}
\end{equation}
where $E_R$ is the recoil energy,
$d\Omega_q=d\phi d\cos\theta$ denotes an infinitesimal solid angle around the recoil direction $\hat{\bf q}= \vec{\bf q}/q$, $q=|\vec{\bf q}|$ is the magnitude of the recoil momentum,  $\mu=mM/(m+M)$ is the reduced WIMP-nucleus mass, $q/2\mu$ is the minimum velocity a WIMP must have to impart a recoil momentum $q$ to the nucleus, $ \rho$ is the dark matter density in the solar neighborhood, $\sigma_0$ is the total scattering cross section of the WIMP  with a (fictitious) point-like nucleus, and $S(q)$ is the nuclear form factor normalized to 1. We concentrate here on WIMPs with spin-independent interactions.

For a truncated Maxwellian WIMP velocity distribution with respect to the Galaxy, with dispersion $\sigma_v$ and truncated at the escape speed $v_{\rm esc}$, the Radon-transform is~\cite{Gondolo:2002}
\begin{equation}
\hat{f}_{\rm lab}\!\left( \frac{q}{2\mu}, \hat{\bf q} \right)=\frac{1}{{N_{\rm esc}(2\pi \sigma_v^2)^{1/2}}}~{\left\{\exp{\left[-\frac{\left[ (q/2\mu) + \hat{\bf q} . {\bf V}_{\rm lab}\right]^2}{2\sigma_v^2}\right]}-\exp{\left[\frac{-v_{\rm esc}^2}{2\sigma_v^2}\right]}\right\}},
\label{fhatTM}
\end{equation}
 if $(q/2\mu) + \hat{\bf q} . {\bf V}_{\rm lab} < v_{\rm esc}$, and zero otherwise, where
\begin{equation}
N_{esc}=\mathop{\rm erf}\left(\frac{v_{\rm esc}}{\sqrt{2}\sigma_v}\right)-\sqrt{\frac{2}{\pi}}\frac{v_{\rm esc}}{\sigma_v}\exp{\left[-\frac{v_{\rm esc}^2}{2\sigma_v^2} \right]}.
\label{Radon-transform}
\end{equation}
Here we are assuming the detector has a velocity $\textbf{V}_{\rm lab}$ with respect to the Galaxy (thus  $- \textbf{V}_{\rm lab}$ is the average velocity of the WIMPs with respect to the detector). ${\bf V}_{\rm lab}$ is defined in terms of the galactic rotation velocity ${\bf V}_{\rm {Gal Rot}}$ at the position of the Sun (or Local Standard of Rest (LSR) velocity), Sun's peculiar velocity ${\bf V}_{\rm {Solar}}$ in the LSR, Earth's translational velocity ${\bf V}_{\rm {Earth Rev}}$ with respect to the Sun, and the  velocity of Earth's rotation around itself ${\bf V}_{\rm {Earth Rot}}$. Thus, ${\bf V}_{\rm {lab}}={\bf V}_{\rm {Gal Rot}}+{\bf V}_{\rm {Solar}}+{\bf V}_{\rm {Earth Rev}}+{\bf V}_{\rm {Earth Rot}}$. We take $V_{\rm GalRot}$ either 220 km/s or 280 km/s, as reasonable low and high values (as done in Ref~\cite{Green-2010}),  which correspond to $V_{\rm lab}$ either 228.4 km/s or 288.3 km/s, respectively. The presence of $\hat{\bf q} . {\bf V}_{\rm lab}$ in Eq.~\ref{fhatTM} means that in order to compute the differential rate we  need  to orient the nuclear recoil direction $\hat{\textbf{q}}$ with respect to ${\bf V}_{\rm lab}$.

The differential energy spectrum is given by
\begin{equation}
\frac{dR}{dE}=\int{\frac{dR}{dE_R d\Omega_q}p(E,E_R,\hat{\textbf{q}})d\Omega_q dE_R},
\label{def-Rate}
\end{equation}
where $p(E,E_R,\hat{\textbf{q}})dE$ is the probability that an energy $E$ is measured when a nucleus recoils in the direction $\hat{\textbf{q}}$ with initial energy $E_R$. With our analytic approach we cannot estimate the importance of dechanneling mechanisms, such as the presence of lattice imperfections, impurities or dopants. Thus we disregard dechanneling, and assume that a recoiling nucleus can only either be channeled, in which case the measured energy is the whole initial recoil energy $E=E_R$ (first term in the following equation) or not channeled, in which case the measured energy is $E= Q E_R$ (second term),
\begin{equation}
p(E,E_R,\hat{\textbf{q}})=\chi(E_R, \hat{\textbf{q}})\delta(E-E_R)+[1-\chi(E_R, \hat{\textbf{q}})]\delta(E-QE_R),
\label{prob}
\end{equation}
where $\chi(E_R, \hat{\textbf{q}})$ is the probability that a nucleus with recoil energy $E_R$ is channeled in a given direction $\hat{\textbf{q}}$.

Inserting Eqs.~\ref{eq: rate} and \ref{prob} in Eq.~\ref{def-Rate}, we find the measured differential rate. The integral in Eq.~\ref{def-Rate} cannot be computed analytically, and we integrate numerically using HEALPix.

\section{Daily Modulation in NaI Crystals}
We present here the daily modulation amplitude expected in NaI crystals, assuming that WIMPs have a truncated Maxwellian velocity distribution with $v_{\rm esc}=650$ km/s and  $\sigma_v=300$ km/s~\cite{Gondolo:2002}.

The spin-independent detection rate of WIMPs has a time dependence through the Radon transform $\hat{ f}_{\rm lab}$. Notice that $\hat{ f}_{\rm lab}$ (see Eq.~\ref{fhatTM}) changes during a day through the $(\hat{\bf q} . {\bf V}_{\rm lab})$ factor appearing in the exponent  and the dependence of ${\bf V}_{\rm lab}$ on ${\bf V}_{\rm {EarthRot}}$.

Here we show  the  signal rate as a function of time during a particular arbitrary Solar day (September 25, 2010). The total rate consists of signal plus background, $R_T = R_s +R_b$, and we assume that there is no daily modulation in the background. We define  the relative  signal modulation amplitude $A_s$ (taking into account the signal only) in terms of the maximum and minimum daily signal rate $R_s$ as
 \begin{equation}
 A_s= \frac{R_{s {\rm -max}}-R_{s{\rm -min}}}{R_{s{\rm-max}}+R_{s{\rm-min}}}.
 \end{equation}
The average rate due to the signal alone is $R_s =(R_{s {\rm -max}}+R_{s {\rm -min}})/2$.

\begin{figure}
\begin{center}
  \includegraphics[height=160pt]{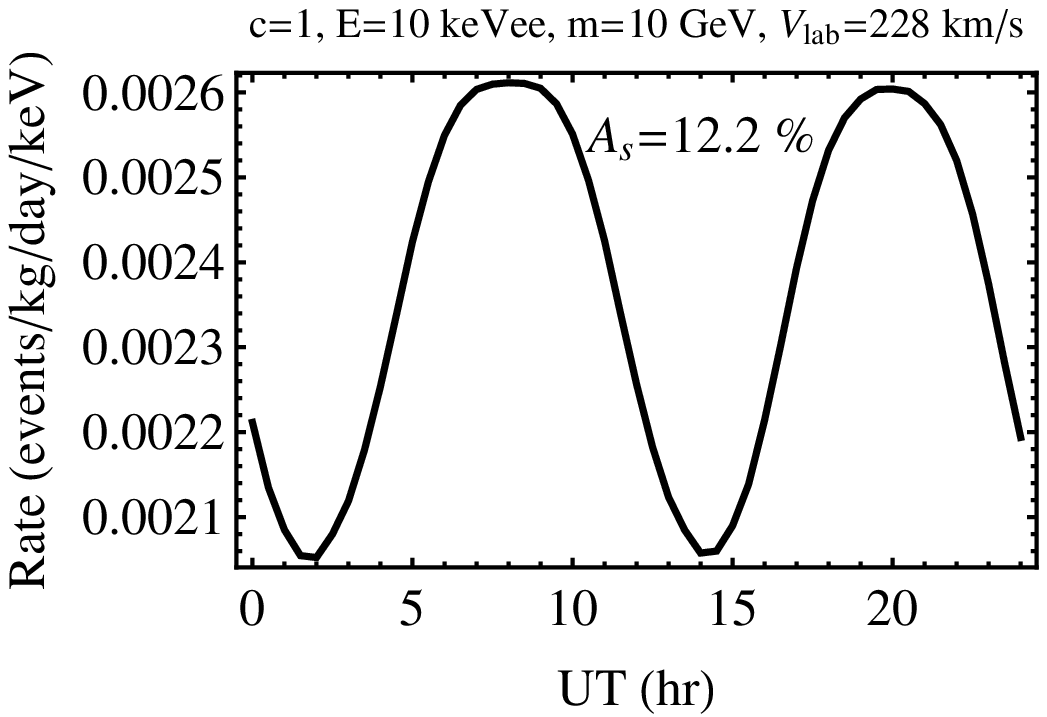}
  \includegraphics[height=160pt]{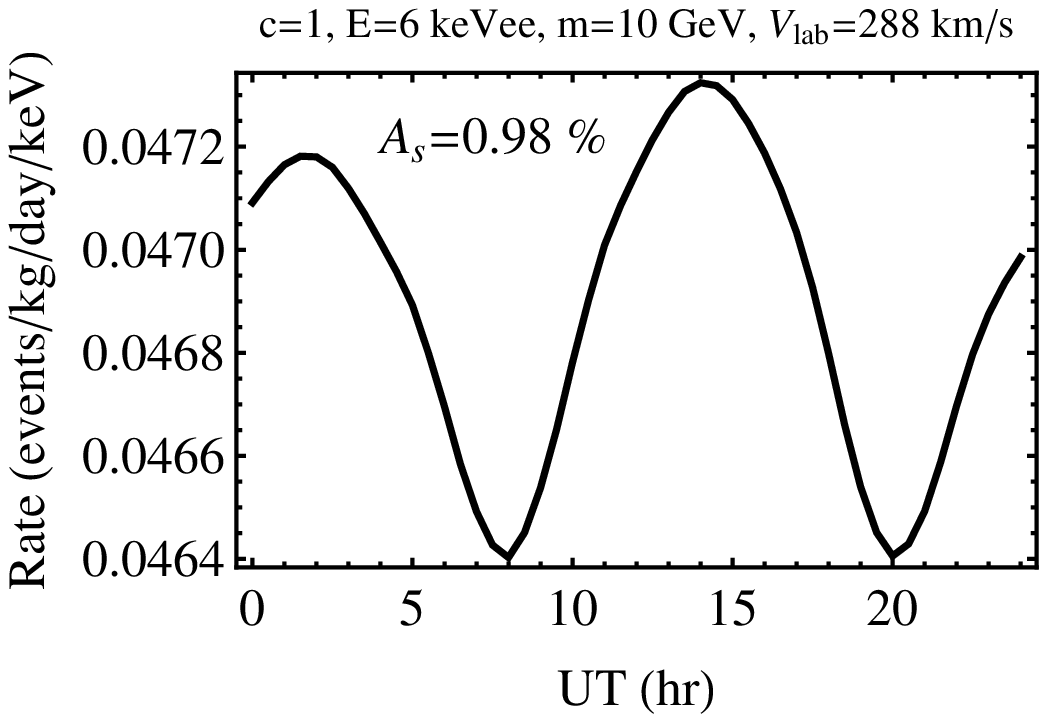}\\
  \caption{Signal rate (in events/kg-day-keVee) as function of the Universal Time (UT) during 24 hours for $m=10$ GeV and (a) $E=10$ keVee and $V_{\rm {lab}}=228.4$ km/s, and (b) $E=6$ keVee  and $V_{\rm {lab}}=288.3$ km/s. The parameters used are $Q_{\rm Na}=0.2$, $Q_{\rm I}=0.09$, $\sigma_p=2 \times 10^{-40} \textrm{cm}^2$, $c=1$ for temperature effects and a crystal temperature of $T=293$ K.}
	\label{RateExample}
\end{center}
\end{figure}

Exploring the parameter space of WIMP mass and WIMP-proton cross section for different recoil energies we find that the relative modulation amplitudes $A_s$ can be large,  even more than 10\%  for some combination of parameters. We show two examples in  Fig.~\ref{RateExample}, where we plot the  signal rate (in events/kg/day/keVee) as a function of  the Universal Time (UT) during 24 hours.  We find that the largest $A_s$ happen when the signal is only due to channeling. This happens when there are no WIMPs in the galactic halo with large enough kinetic energy to provide the observed energy if the recoil is not channeled.

The detectability  of a particular amplitude of daily modulation depends on the exposure and background of a particular experiment. The former DAMA/NaI and the DAMA/LIBRA experiments have a very large cumulative exposure, 1.17 ton $\times$ year. However even with this large exposure, we find that the daily modulations we predict are not observable.  To observe the daily modulation,  the  total number of events  $N_T$  ($N_s$ signal plus $N_b$ background events) over the duration of the experiment  should be divided into two bins,  the ``high-rate''  bin with $N_{T {\rm -max}}$ events and the ``low-rate'' bin with $N_{T {\rm -min}}$ events,  so that  $N_T=N_{T {\rm -max}}+N_{T {\rm -min}}$.  For the daily modulation to be observable at the 3$\sigma$ level one should have
\begin{equation}
N_{T {\rm -max}}-N_{T {\rm -min}}= A_s N_s > 3\sigma \simeq 3 \sqrt{N_T/2},
\label{RateCond}
\end{equation}
where  $\sigma^2 \simeq N_T/2$ because, with a small modulation, on average $N_{T {\rm -max}} \simeq N_{T {\rm -min}} \simeq N_T/2$.

If the detector exposure is $M T$ in kg-day and we take bins of width $\Delta E$  in keVee, then $N_T =R_T M T~ \Delta E$ and $N_s =R_s M T \Delta E$,  where the rates are in events/kg-day-keVee. Thus the condition in Eq.~\ref{RateCond} becomes
\begin{equation}
R_s^2/R_T > 9 /(2 A_s^2 M T~ \Delta E).
\label{RateCond2}
\end{equation}
The total rate of the DAMA experiment at low energies between 4 keVee and 10 keVee is $R_T \simeq 1 ~{\rm events/kg/day/keVee}$~\cite{DAMA-bckg}. This rate is much larger than the signal rates we predict  and is, therefore, dominated by background. We choose here an energy bin $\Delta E \simeq 1$ keVee, narrow enough to assume the signal rate to be constant in it and compatible with the energy resolution of DAMA~\cite{Bernabei:2008}. With the cumulative exposure of DAMA, the condition in Eq.~\ref{RateCond2} for relative daily modulation amplitude $A_s$ observable at 3$\sigma$ is $R_s~A_s>3.2 \times 10^{-3}~{\rm events/kg/day/keVee}$ or
\begin{equation}
R_{s {\rm -max}}-R_{s {\rm -min}} > 6.4 \times 10^{-3}~{\rm events/kg/day/keVee}.
\label{RateCond4}
\end{equation}
Even the largest relative daily modulations we find, shown in Fig.~\ref{RateExample}, are not observable in the DAMA data according to Eq.~\ref{RateCond4}.

One could ask what exposure would be needed with the current total rate in the DAMA experiment  to make the daily modulation observable. Setting $R_T \simeq 1$ events/kg/day/keVee in Eq.~\ref{RateCond2}, we obtain
\begin{equation}
\frac{M T \Delta E}{{\rm (events/kg/day/keVee)}} > \frac{9}{2 \left(A_s~R_s \right)^2}=  \frac{18}{\left(R_{s {\rm -max}}-R_{s {\rm -min}}\right)^2}  .
\end{equation}
For the case with the highest rate difference we found  ($m=10$ GeV, $E=6$ keVee and $V_{\rm lab}=288.3$ km/s, shown in Fig.~\ref{RateExample}.b) and with $\Delta E \simeq$ 1 keVee we would require an exposure 40 times larger than the current exposure of DAMA.

\section{Future Prospects for Other Experiments}
 We have computed the daily modulation due to channeling in other material such as Ge, solid Xe and solid Ne, and we find that  it will be very difficult to observe.  For light WIMPs the cross section can be larger than for heavier ones without violating experimental bounds,  $\sigma_p=10^{-39} \textrm{cm}^2$~\cite{TEXONO:2009, Aalseth:2008rx} and this favors the detection of the daily modulation.  We find that  for a WIMP mass $m=5$ GeV the daily modulation due to channeling may be observable in solid Ne if the signal would be above threshold and  assuming no background. For example for a solid Ne detector operating at 23 K at Gran Sasso, for $E=10$ keV, assuming $Q_{\rm Ne} =0.25$~\cite{Tretyak}, $c=1$ and $V_{\rm {lab}}=228.4$ km/s, we find that the exposure needed to observe the modulation at 3$\sigma$ is $MT=0.33$ ton year. The usual rate is zero in this case, and the modulation is just due to channeling.

The daily modulation in the usual rate may be possible to observe in future non-directional detectors with negligible background. The required exposures to have an observable daily modulation at the 3$\sigma$ level in future solid or liquid detectors are shown in Table~\ref{table:Observability} for cases in which channeling is negligible (or absent in liquid detectors), and the daily modulation is due purely to the change of WIMP momentum distribution in the lab frame as the Earth rotates around itself. The exposures are calculated assuming no background and with $V_{\rm lab}=228.4$ km/s for a Ge detector at the Soudan Underground Laboratory (with $Q_{\rm Ge} =0.2$~\cite{TEXONO:2009}) and all other detectors at Gran Sasso.

\begin{table}[h]
\begin{center}
\caption{Observability in Future Detectors}
\begin{tabular}{|l|c|c|c|c|}
\hline \textbf{Detector} & \textbf{$E_R$ (keV)} & \textbf{$m$ (GeV)} &
\textbf{$\sigma_p$ ($\textrm{cm}^2$)}  & \textbf{$MT$ (ton year)}
\\
\hline Ge & 10 & 7 & $10^{-41}$ & 11 \\
\hline Solid Ne & 8 & 5 & $10^{-39}$ & 0.8 \\
\hline Liquid Ne & 14 & 5 & $10^{-39}$ & 3.4 \\
\hline Liquid Xe & 3 & 5 & $10^{-39}$ & 0.09 \\
\hline Liquid Ar & 8 & 5 & $10^{-39}$ & 0.9 \\
\hline
\end{tabular}
\label{table:Observability}
\end{center}
\end{table}

\section{Conclusions}

We have studied the channeling of ions recoiling after collisions with WIMPs in different crystalline detectors such as NaI (Tl), Si and Ge. Channeled ions move within the crystal along symmetry axes and planes and suffer a series of small-angle scatterings  that maintain them in the open ``channels''  between the rows or planes of lattice atoms and thus penetrate much further into the crystal than in other directions. Ions which start their motion close to the center of a channel, at an initial angle $\psi$, are channeled if the initial angle  is smaller than a critical angle, and are not  channeled otherwise.  With the simple model of dechanneling we used for NaI, we could reproduce the channeling fractions computed by the DAMA collaboration. However this is not the case in direct dark matter detectors, since the recoiling ions start their motion at or close to their original lattice sites. For recoiling ions, blocking effects (neglected in the DAMA calculation) are important, and the channeling fraction is smaller. As argued originally by Lindhard, in a perfect lattice and in the absence of energy-loss processes, the probability that a particle starting from a lattice site is channeled would be zero. However, due to vibrations  in the crystal, the atom that interacts with a WIMP may be displaced from its position in a perfect lattice, and there is a non-zero probability of channeling. As seen in Fig.~\ref{FracNaI-Final} for NaI, without including dechanneling, the channeling fraction is never larger than 5\%.

We have also studied the possibility of a daily modulation due to channeling or in the usual non-channeled rate, which could be a background free signature of dark matter. We find large daily modulation amplitudes for NaI (even more than 10\%) which are not observable in DAMA. However, the daily modulation might be detectable in other experiments such as Ge, solid and liquid Ne, and liquid Xe and Ar with smaller background or larger exposure. We intend to further explore the observability of a daily modulation in future experiments for different halo models in future work.



\begin{acknowledgments}
N.B. and Graciela Gelmini were supported in part by the US Department of Energy Grant
DE-FG03-91ER40662, Task C. Paolo Gondolo was supported in part by  the NFS
grant PHY-0756962 at the University of Utah.
\end{acknowledgments}

\bigskip 

\end{document}